\documentstyle[aps]{revtex}

\title{Chiral Symmetry Breaking in the weakly coupled QED in a 
Magnetic Field\thanks{This work is done in collaboration with 
V.P.~Gusynin and V.A.~Miransky [1,2,3]}}

\author{Igor A.~Shovkovy}

\address{Bogolyubov Institute for Theoretical Physics,
252143, Kiev, Ukraine}

\draft
\begin{document}
\maketitle

\begin{abstract}
The catalysis of chiral symmetry breaking in the 
massless weakly coupled QED in a magnetic field is studied. It 
is shown that the effect is due to the dimensional reduction 
$D\to D-2$ in the dynamics of fermion pairing in a magnetic 
field. The dynamical mass of fermions (energy gap in the fermion 
spectrum) is determined. The temperature of the symmetry 
restoration is estimated analytically.  
\end{abstract} 


Recently a new nonperturbative phenomenon, the catalysis of chiral 
symmetry breaking in an external magnetic field, was revealed in 
the massless weakly coupled QED \cite{prdQED,npb}.  The key 
observation which led to the new effect in QED came from the much 
simpler NJL model \cite{prl,plb,prdNJL}. As was noticed there the 
catalysis induced by the magnetic field should have a universal 
(model-independent) character. The roots of the universality lie in 
the effective dimensional reduction $D\rightarrow D-2$ in the 
fermion pairing which results from the dominance of the lowest 
Landau level (LLL) in the low-energy dynamics. As concrete models, 
the Nambu-Jona-Lasinio (NJL) model as well as QED in 2+1 and 3+1 
dimensions were considered 
\cite{prdQED,npb,prl,plb,prdNJL,Klim,Ng1,Hong,QED3}.  The effect of 
catalyzing the dynamical chiral symmetry breaking under the 
influence of a magnetic field was extended also to the case of 
external non-abelian chromomagnetic fields \cite{Igor,Ebert} and 
finite temperatures \cite{qedt_c,Smilga,Ng2}, as well as to the 
supersymmetric NJL model \cite{Elias}.

Here I shall briefly review the analysis of the effect in the 
weakly coupled QED$_4$ \cite{prdQED,npb} at zero temperature and 
then consider the question of the chiral symmetry restoration at 
finite temperature \cite{qedt_c}. 

The Lagrangian density of massless QED$_4$ in a magnetic field is
\begin{equation}
{\cal L}= -{1\over 4}F^{\mu\nu}F_{\mu\nu}
+ {1\over2}[\bar\psi,i\gamma^\mu D_\mu\psi],
\label{density}
\end{equation}
where the covariant derivative $D_\mu $ is
\begin{eqnarray}
D_\mu=\partial_\mu-ie(A_\mu^{ext}+A_\mu),\quad
A_\mu^{ext}=\left(0,-{B\over2}x_2,{B\over2}x_1,0\right).
\end{eqnarray}
Besides the Dirac index ($n$), the fermion field carries an 
additional flavor index $a=1,2,\ldots,N_f$. The Lagrangian density 
(\ref{density}) is invariant under the chiral $SU_{\rm 
L}(N_f)\times SU_{\rm R}(N_f)\times U_{\rm L+R}(1)$ (here I do not 
discuss the anomaly connected with the current $j_{5\mu}$).  As is 
known, there is no spontaneous chiral symmetry breaking at $B=0$ in 
the weak coupling phase of QED \cite{RNC}.  However, the magnetic 
field changes the situation drastically: at $B\neq0$ the chiral 
symmetry is broken down to $SU_{\rm V}(N_f) \equiv SU_{\rm L+R}(N_f)$
and there appear $N_f^2-1$  gapless Nambu-Goldstone (NG) 
bosons composed from fermion-antifermion pairs. The dynamical mass 
(energy gap) for a fermion can be defined by considering the 
Bethe-Salpeter (BS) equation for NG boson \cite{prdQED,npb} or the 
Schwinger-Dyson (SD) equation for the dynamical mass function 
\cite{Ng1,Hong}. Here I follow the first approach.

The homogeneous BS equation for the $N_f^2-1$ NG bound states takes 
the form \cite{Mir}:
\begin{eqnarray}
\chi^\beta_{AB}(x,y;P) &=& -i \int d^4x_1 d^4y_1
d^4x_2d^4y_2 G_{AA_1}(x,x_1) 
K_{A_1B_1;A_2B_2}(x_1y_1,x_2y_2) \chi^\beta_{A_2B_2}(x_2,y_2;P)
G_{B_1B}(y_2,y)~,    \label{94}
\end{eqnarray}
where the BS wave function $\chi^\beta_{AB}=
\langle0|T\psi_A(x)\bar\psi_B(y)|P;\beta\rangle$, $\beta=1,\ldots,
N_f^2-1$, and the fermion propagator $G_{AB}(x,y)=
\langle0|T\psi_A(x)\bar\psi_B(y)|0\rangle$; the indices
$A=(na)$ and $B=(mb)$ include both Dirac $(n,m)$ and flavor
$(a,b)$ indices.

The BS kernel in leading order in $\alpha$ is \cite{Mir}:
\begin{eqnarray}
K_{A_1B_1;A_2B_2}(x_1y_1,x_2,y_2)&=&
-4\pi i\alpha \delta_{a_1a_2}\delta_{b_2b_1}\gamma^\mu_{n_1n_2}
\gamma^\nu_{m_2m_1}{\cal{D}}_{\mu\nu}(y_2-x_2) 
\delta(x_1-x_2)\delta(y_1-y_2)\nonumber
\\&&+4\pi i\alpha\delta_{a_1b_1}
\delta_{b_2a_2}\gamma^\mu_{n_1m_1}\gamma^\nu_{m_2n_2}
{\cal{D}}_{\mu\nu}(x_1-x_2) \delta(x_1-y_1)\delta(x_2-y_2),   
\label{95}
\end{eqnarray}
where the photon propagator is
\begin{eqnarray}
{\cal{D}}_{\mu\nu}(x)={-i\over (2\pi)^4} \int d^4k e^{ikx}
\Bigg(g_{\mu\nu}-\lambda {k_\mu k_\nu \over k^2}\Bigg)
{1\over k^2}                         
\label{96}
\end{eqnarray}
($\lambda$ is a gauge parameter).  The first term on the 
right--hand side of Eq.(\ref{95}) corresponds to the ladder 
approximation.  The second (annihilation) term does not contribute 
to the BS equation for NG bosons. Then the equation takes the 
form:
\begin{eqnarray}
\chi^\beta_{AB}(x,y;P)&=&-4\pi\alpha \int d^4x_1 d^4y_1
S_{AA_1}(x,x_1) \delta_{a_1a_2} \gamma^\mu_{n_1n_2}
\chi^\beta_{A_2B_2}(x_1,y_1;P) 
\delta_{b_2b_1}\gamma^\nu_{m_2m_1}S_{B_1B}(y_1,y)
{\cal{D}}_{\mu\nu}(y_1-x_1)~, \label{97}
\end{eqnarray}
where, since the lowest in $\alpha$ (ladder) approximation is 
used, the full fermion propagator $G_{AB}(x,y)$ is replaced by the
propagator $S$ of a free fermion (with the mass $m=m_{\rm dyn}$) in
a magnetic field.

Using the new variables, the center of mass coordinate 
$R=(x+y)/2$, and the relative coordinate $r=x-y$, 
Eq.(\ref{97}) can be rewritten as 
\begin{eqnarray} 
&&\tilde\chi_{nm}(R,r;P)=-4\pi\alpha \int d^4R_1d^4r_1 
\tilde S_{nn_1}\Bigg(R-R_1+{r-r_1\over 2}\Bigg)
\gamma^\mu_{n_1n_2}\tilde\chi_{n_2m_2}(R_1,r_1;P)
\gamma^\nu_{m_2m_1} 
\nonumber\\
&&\tilde S_{m_1m}\Bigg({r-r_1\over 2}-R+R_1\Bigg)
{\cal{D}}_{\mu\nu}(-r_1)
\exp\bigl[-ie(r+r_1)^\mu A_\mu^{\rm ext}(R-R_1)\bigr]  
\times \exp\bigl[iP(R-R_1)\bigr]~.  \label{98}
\end{eqnarray}
Here the function $\tilde\chi_{nm}(R,r;P)$ is defined from the 
equation
\begin{eqnarray}
\chi^\beta_{AB}(x,y;P) \equiv
\langle0|T\psi_A(x)\bar\psi_B(y)|P,\beta\rangle
=\lambda^\beta_{ab}e^{-iPR} \exp\bigl[ ier^\mu A_\mu^
{\rm ext}(R)\bigr]\tilde\chi_{nm}(R,r;P)&& \label{99}
\end{eqnarray}
where $\lambda^\beta$ are $N_f^2-1$ flavor matrices (${\rm 
tr}(\lambda^\beta \lambda^\gamma)=2\delta^{\beta\gamma};~ 
\beta,\gamma \equiv 1,\ldots,N_f^2-1$).  The important fact is that 
the only effect of translation symmetry breaking by the magnetic 
field is given by the Schwinger phase factor in Eq.(\ref{99}). 
Thus, Eq.(\ref{98}) admits the "translation invariant" solution 
$\tilde\chi_{nm}(R,r;P)=\tilde\chi(r;P)$.  In momentum space, one 
obtain
\begin{eqnarray}
&&\tilde\chi_{nm}(p;P)=-4\pi\alpha \int
{d^2q_{\perp} d^2R_{\perp} d^2k_{\perp} d^2k_{\parallel}\over
(2\pi)^6} \exp\bigl[i({\bf P}_{\perp}
-{\bf q}_{\perp}){\bf R}_{\perp}\bigr]
\tilde S_{nn_1}\Bigg(p_{\parallel}
+{P_{\parallel}\over 2},{\bf p}_{\perp}
+e{\bf A}^{\rm ext}({\bf R}_{\perp})
+{{\bf q}_{\perp}\over 2}\Bigg) \nonumber\\
&\times& \gamma^\mu_{n_1n_2}\tilde\chi_{n_2m_2}(k,P)
\gamma^\nu_{m_2m_1}\tilde S_{m_1m}
\Bigg(p_{\parallel}-{P_{\parallel}\over 2},
{\bf p}_{\perp}+e{\bf A}^{\rm ext}({\bf R}_{\perp})-
{{\bf q}_{\perp}\over 2}\Bigg)
{\cal{D}}_{\mu\nu}\bigl(k_{\parallel}-p_{\parallel},
{\bf k}_{\perp}-{\bf p}_{\perp}
-2e{\bf A}^{\rm ext}({\bf R}_{\perp})\bigr) \label{100}
\end{eqnarray}
(recall that $p_{\parallel}\equiv (p^0,p^4)$, ${\bf p}_{\perp} 
\equiv (p^1,p^2)$). Henceforth I shall consider the equation with 
the total momentum $P_\mu\to 0$.

The crucial point for further analysis will be the assumption that 
$m_{\rm dyn} \ll \sqrt{|eB|}$ and that the region mostly 
responsible for generating the mass is the infrared region with $k 
\alt \sqrt{|eB|}$.  As it will be seen, this assumption is 
self-consistent.  The assumption allows to replace the fermion 
propagator $\tilde S_{nm}$ by the pole contribution of the LLL:  
\begin{eqnarray} 
\tilde S^{(0)}(k)=i~\exp\left(-{{\bf k}_{\perp}^2\over |eB|}\right) 
{k^0\gamma^0-k^3\gamma^3+m_{\rm dyn}\over k_0^2-k_3^2-m_{\rm dyn}^2}
\left(1-i\gamma^1\gamma^2 {\rm sign}(eB)\right)~,
\label{LLL}
\end{eqnarray}
and Eq.(\ref{100}) transforms into the following one:
\begin{eqnarray}
\rho(p_{\parallel},{\bf p}_{\perp}) &=&
{2\alpha\ell^2\over (2\pi)^4} e^{-\ell^2{\bf p}_{\perp}^2} \int
d^2A_{\perp} d^2k_{\perp} d^2k_{\parallel}
e^{-\ell^2{\bf A}_{\perp}^2}(1-i\gamma^1\gamma^2)\gamma^\mu 
\nonumber\\
&&\times {\hat k_{\parallel}+m_{\rm dyn}\over
k^2_{\parallel}-m^2_{\rm dyn}}
\rho(k_{\parallel},{\bf k}_{\perp})
{\hat k_{\parallel}+m_{\rm dyn}\over k^2_{\parallel}-m^2_{\rm dyn}}
\gamma^\nu(1-i\gamma^1\gamma^2) 
{\cal{D}}_{\mu\nu}(k_{\parallel}-p_{\parallel},
{\bf k}_{\perp}-{\bf A}_{\perp})~,     \label{101}
\end{eqnarray}
where $\rho(p_{\parallel}, {\bf p}_{\perp})= (\hat p_{\parallel} 
-m_{\rm dyn}) \tilde\chi(p) (\hat p_{\parallel} -m_{\rm dyn})$ and 
$\ell=1/\sqrt{|eB|}$ is the magnetic field length.  
Equation (\ref{101}) implies that $\rho(p_{\parallel},
{\bf p}_{\perp}) =\exp(-\ell^2 {\bf p}_{\perp}^2) 
\varphi(p_{\parallel})$, where $\varphi(p_{\parallel})$ 
satisfies the equation 
\begin{eqnarray} 
\varphi(p_{\parallel}) &=&
{\pi\alpha\over (2\pi)^4} \int d^2k_{\parallel}
(1-i\gamma^1\gamma^2)\gamma^\mu
{\hat k_{\parallel}+m_{\rm dyn}\over k^2_{\parallel}-m^2_{\rm dyn}}
\varphi(k_{\parallel}) 
{\hat k_{\parallel}+m_{\rm dyn}\over k^2_{\parallel}-
m^2_{\rm dyn}} \gamma^\nu(1-i\gamma^1  \gamma^2) 
D^{\parallel}_{\mu\nu}
(k_{\parallel}-p_{\parallel})~. \label{102}
\end{eqnarray}
  Here
\begin{eqnarray}
D^{\parallel}_{\mu\nu}(k_{\parallel}-p_{\parallel})=
\int d^2k_{\perp} \exp\Bigg(-{\ell^2 {\bf k}^2_{\perp}\over 2}\Bigg)
{\cal{D}}_{\mu\nu}(k_{\parallel}-p_{\parallel},{\bf k}_{\perp})~.
\label{103}
\end{eqnarray}
Thus, the BS equation has been reduced to a two--dimensional
integral equation.  

Henceforth I shall use Euclidean space with $k_4=-ik^0$.  Then, 
because of the symmetry $SO(2)\times SO(2)\times {\cal{P}}$ in a 
magnetic field, one arrives at the following matrix structure for 
$\varphi(p_{\parallel})$:
\begin{eqnarray}
\varphi(p_{\parallel})=\gamma_5(A+i\gamma_1\gamma_2 B+
\hat p_{\parallel}C+i\gamma_1\gamma_2 \hat p_{\parallel} D)
\label{104a}
\end{eqnarray}
where $A,B,C$ and $D$ are functions of $p^2_{\parallel}$.

Let me begin the analysis of Eq.(\ref{102}) by choosing the 
Feynman gauge. Then 
\begin{eqnarray} 
D^{\parallel}_{\mu\nu}(k_{\parallel}-p_{\parallel})=
i\delta_{\mu\nu} \pi \int^\infty_0
{dx \exp(-\ell^2 x/2)\over (k_{\parallel}-p_{\parallel})^2+x}~,
\label{104}
\end{eqnarray}
and, substituting the expression (\ref{104a}) for 
$\varphi(p_{\parallel})$ into Eq.(\ref{102}), one finds that 
$B=-A$, $C=D=0$, and the function $A(p_{\parallel})$ satisfies 
(henceforth I omit the subscript $\parallel$ in momenta) the 
equation:
\begin{eqnarray}
A(p)={\alpha\over 2\pi^2}\int {d^2kA(k)\over k^2+m^2_{\rm dyn}}
\int^\infty_0
{dx \exp(-\ell^2x/2)\over ({\bf k}-{\bf p})^2+x}. \label{106}
\end{eqnarray}
This equation has been analyzed in \cite{npb,Hong}. Particularly, 
as was shown in \cite{npb} (see Appendix C), in the case of weak 
coupling $\alpha$, the mass function $A(p)$ remains almost constant 
in the range of momenta $0<p^2\alt 1/\ell^2$ and decays like 
$1/p^2$ outside that region.  To get an estimate for $m_{\rm dyn}$ 
at $\alpha\ll 1$, the external momentum is set to be zero. Then the 
main contribution of the integral is formed in the infrared region 
with $k^2\alt 1/\ell^2$. The latter validates in its turn the 
substitution $A(k) \rightarrow A(0)$ in the integrand of 
(\ref{106}), and finally one comes to the following gap equation 
\begin{eqnarray} 
A(0)\simeq\frac{\alpha}{2\pi^2}A(0)
\int\frac{d^2k}{k^2+m^2_{\rm dyn}}
\int_0^\infty\frac{dx\exp(-x\ell^2/2)}{k^2+x},
\label{approx}
\end{eqnarray}
i.e.
\begin{eqnarray}
1\simeq\frac{\alpha}{2\pi}\int_0^\infty
\frac{dx\exp(-m^2_{\rm dyn}\ell^2x/2)}{x-1}\log x
\simeq \frac{\alpha}{4\pi}\log^2
\left(\frac{m^2_{\rm dyn}\ell^2}{2}\right).
\label{gapeq}
\end{eqnarray}
The latter leads to the following expression for the 
dynamical mass:
\begin{equation}
m_{\rm dyn}\simeq C\sqrt{|eB|}\exp
\left[-\sqrt{\pi\over\alpha}\right],
\label{massdyn}
\end{equation}
where $C$ is a constant of order one. The exponential factor 
displays the nonperturbative nature of this result. 

In order to avoid any confusion let me mention that the electron 
self-energy (under the influence of a strong external magnetic 
field) in the standard perturbation theory \cite{Jancov,Loskutov} 
also reveals the double logarithmic asymptotics. However, the 
chiral limit which of the prime interest here is unreachable in the 
framework of the perturbation theory. The latter explains why the 
results known for several decades did not lead to what is called 
here the catalysis of chiral symmetry breaking by a magnetic field.

More accurate analysis which takes into account the momentum 
dependence of the mass function leads to the result \cite{npb} 
\begin{equation}
m_{\rm dyn}\simeq C\sqrt{|eB|}\exp\left[-{\pi\over2}
\sqrt{\frac{\pi}{2\alpha}}\right]. \label{solution}
\end{equation}
Notice that the ratio of the powers of this exponent and that in 
Eq.(\ref {massdyn}) is $\pi/2\sqrt{2}\simeq 1.1$, thus the 
approximation used above is rather reliable.

It is worth to note that Eq.(\ref{106}) reduces to the 
two--dimensional (another manifestation of the effective 
dimensional reduction in the dynamics of fermion pairing) 
Schr\"odinger equation
\begin{eqnarray}
\bigl(-\Delta+m^2_{\rm dyn}+V({\bf r})\bigr) \Psi({\bf r})=0~,
\label{108}
\end{eqnarray} 
where the wave function $\Psi({\bf r})$ is defined as follows
\begin{eqnarray}
\Psi({\bf r}) = \int {d^2k\over (2\pi)^2}
{A(k)\over k^2+m^2_{\rm dyn}} e^{i{\bf kr}}~, \label{107}
\end{eqnarray}
and the potential $V({\bf r})$ reads
\begin{eqnarray}
V({\bf r})&=&{\alpha\over \pi\ell^2} 
\exp\Bigg({r^2\over 2\ell^2}\Bigg)
Ei\Bigg(-{r^2\over 2\ell^2}\Bigg) \label{109}
\end{eqnarray}
Note that this potential has the asymptotics $V({\bf r}) \simeq 
-2\alpha/\pi r^2$ as $r\to\infty$.  The Schr\"odinger equation with 
such potentials was studied in \cite{PerPop}. Their result for 
$E(\alpha)$ agrees with the solution in (\ref{solution}).  

As is known, the ladder approximation is not gauge invariant.  
However, the consideration of the general covariant gauge 
\cite{npb} shows that the leading term in $\ln(m_{\rm 
dyn}^2\ell^2)$, namely $\ln(m_{\rm dyn}^2\ell^2) \simeq 
-\pi\sqrt{\pi/2\alpha}$, is the same in all covariant gauges.  
Thus, in spite of absence of the gauge invariance in the ladder 
approximation, there are no doubts about the existence of the 
effect.

Up to now I have considered the ladder approximation in QED in a 
magnetic field. Let me say several words about higher order 
contributions (such as vacuum polarization).  As is shown in 
\cite{npb}, taking into account the vacuum polarization results in 
an equation similar to (\ref{106}) but with $\alpha$ replaced by 
$\alpha/2$.  Thus, despite the smallness of $\alpha$ the expansion 
in $\alpha$ is broken in the infrared region in the model.  It is a 
challenge to define the class of all relevant diagrams in QED in a 
magnetic field.  Since the QED coupling constant is weak in the 
infrared region, this problem, though hard, seems not to be 
hopeless.

Now let me turn to the case of finite temperature. For introducing 
the Matsubara technique \cite{Abr}, one have to change all the 
expressions in momentum representation according to the rule:  
\begin{eqnarray} 
\int\frac{d^4k}{(2\pi)^4} &\to & iT\sum^{+\infty}_{n=-\infty} 
\int\frac{d^3k}{(2\pi)^3}, \\
k^0 \to  i\omega_n; &\quad & \omega_n=\pi T(2n+1).
\end{eqnarray}
Thus, the analogue of the equation (\ref{106}) (with 
$m^2(T)$ in the denominator instead of $m^2_{\rm dyn}$)
reads
\begin{eqnarray}
A(\omega_{n'},p)=\frac{\alpha}{\pi}
T\sum_{n=-\infty}^\infty \int\limits_{-\infty}^{\infty}
\frac{dk A(\omega_{n},k)} {\omega^2_{n}+k^2+m^2(T)}
\int\limits^\infty_0
\frac{dx\exp(-x\ell^2/2)}{(\omega_{n}-\omega_{n'})^2+(k-p)^2+x},
\label{eq:our1}
\end{eqnarray}
Now if one takes $n'=0, p=0$ in the left hand side of 
Eq.(\ref{eq:our1}) and puts $A (\omega_{n},k) \approx 
A(\omega_{0},0) = const$ in the integrand, he obtains the 
equation
\begin{eqnarray}
1&=&\frac{\alpha}{\pi}T\sum_{n=-\infty}^\infty
\int\limits_{-\infty}^{\infty}
\frac{dk}{\omega_n^2+k^2+m^2(T)} \int\limits^\infty_0
\frac{dx\exp(-x\ell^2/2)}{(\omega_n-\omega_{0})^2+k^2+x}.
\label{eq:our2}
\end{eqnarray}
It is easy to check that the gap equation (\ref{eq:our2})
coincides with the equation (58) in \cite{Ng2}.

The sum in (\ref{eq:our2}) is easily performed:
\begin{eqnarray}
&&1=\frac{\alpha}{\pi} \int\limits_{0}^{\infty}
\int\limits^\infty_0
\frac{dk dx \exp[-x\ell^2/2]}{[(\pi T)^2 +x -m^2(T)]^2 
+(2\pi T)^2 (k^2+m^2(T))}
\nonumber      \\
&&\times\left\{
\frac{(\pi T)^2 +x -m^2(T)}{\sqrt{k^2 +m^2(T)}}
\tanh \left(\frac{\sqrt{k^2+m^2(T)}}{2T}\right)
+\frac{(\pi T)^2 +m^2(T) -x}{\sqrt{k^2+x}}
\coth \left(\frac{\sqrt{k^2+x}}{2T}\right)
\right\}.\label{eq:our3}
\end{eqnarray}
In the limit $T\to0$, Eq.(\ref{eq:our3}) reduces to the 
following one
\begin{eqnarray}
&&1=\frac{\alpha}{\pi} \int\limits_{0}^{\infty}
\int\limits^\infty_0
\frac{dk dx \exp[-x\ell^2/2]}{\sqrt{k^2+x}\sqrt{k^2+m^2_{\rm dyn}}
\left(\sqrt{k^2+x}+\sqrt{k^2+m^2_{\rm dyn}}\right)},
\label{T=0}
\end{eqnarray}
which is just what one obtains from Eq.(\ref{approx}) after
performing the integration over $k_4=-ik_0$.

The equation for the critical temperature is obtained from
(\ref{eq:our3}) by putting $m(T_c)=0$:
\begin{eqnarray}
1=\frac{\alpha}{\pi}
\int\limits_{0}^{\infty} \int\limits^\infty_0
\frac{dk dx e^{-2x(\pi T_c\ell)^2}}{[1/4 +x]^2 +k^2}
\left\{ \frac{1/4 +x}{k} \tanh \left(\pi k\right)
+\frac{1/4 -x}{\sqrt{k^2+x}} \coth \left(\pi\sqrt{k^2+x}\right)
\right\},
\label{eq:t_c}
\end{eqnarray}
where the change of the variables $x\to (2\pi T_{c})^2 x$ and 
$k\to 2\pi T_{c} k$ was made.

By assuming smallness of the critical temperature in comparison 
with the scale put by the magnetic field, $T_c\ell\ll 1$, it is 
seen that the double logarithmic in field contribution in 
Eq.(\ref{eq:t_c}) comes from the region $0< x \alt 1/2(\pi 
T_{c}\ell)^2 $, $1/\pi \alt k<\infty$.  Simple estimate gives:  
\begin{eqnarray}
1\simeq\frac{\alpha}{\pi} \int\limits_0^{1/2(\pi T_{c}\ell)^2} 
dx \int\limits_{1/\pi}^{\infty} \frac{dk}{[1/4 +x]^2 +k^2}\left[
\frac{1/4 +x}{k}+\frac{1/4-x}{\sqrt{k^2+x}}\right]
\simeq \frac{\alpha}{4\pi}\log^2
\left[\frac{1}{2(\pi T_c\ell)^2}\right].&&
\label{integ}
\end{eqnarray}
Then, for the critical temperature, one obtains:
\begin{eqnarray}
T_{c}\approx \sqrt{|eB|}\exp\left[-\sqrt{\frac{\pi}{\alpha}}\right]
\approx m_{\rm dyn}(T=0),
\label{t_c-m_dyn}
\end{eqnarray}
where $m_{\rm dyn}$ is given by (\ref{massdyn}). The relationship 
$T_c\approx m_{\rm dyn}$ between the critical temperature and the 
zero temperature fermion mass was obtained also in NJL model in 
(2+1)- and (3+1)-dimensions \cite{prdNJL,Ebert}.

In passing, let me just briefly note that the photon thermal mass, 
which is of the order of $\sqrt{\alpha} T$ \cite{Weld}, cannot 
change our result for the critical temperature (unlike the vacuum 
polarization). As is easy to check, it would lead to 
the shift in $x$ for a constant of the order of $\alpha$ in the 
integrand of (\ref{integ}).  However, such a shift is absolutely 
irrelevant for our estimate (\ref{t_c-m_dyn}). 

In conclusion, the main result of this paper is the analytic 
estimate for the dynamical mass and the temperature of the chiral 
symmetry restoration in the weakly interacting QED in a background 
magnetic field.  

\acknowledgments{I would like to thank the organizers of both the 
International Workshop "Mathematical Physics - today, Priority 
Technologies -- for tomorrow" and the International School of 
Subnuclear Physics "Highlights: 50 Years Later" for giving me an 
opportunity to present this talk. I am also grateful to my 
collaborators V.P.~Gusynin and V.A.~Miransky.

This study is supported by Foundation of Fundamental Researches of 
Ministry of Sciences of the Ukraine under grant No 2.5.1/003.}

\end{document}